\documentclass{sag00_pdflatex}
\Received{2023 October 31}
\Accepted{2023 December 5}
\Published{2023 December 28}
\SetRunningHead{M. Takayama}{Period modulations of the LSPs}

\usepackage{times}

\renewcommand{\arraystretch}{1.5}

\begin{document}
\SetVolumeData{2023}{6}{6}


\title{Period modulations of the long secondary periods on the AGB stars}
\author{Masaki Takayama\altaffilmark{1}$^{*}$}
\altaffiltext{1}{Nishi-Harima Astronomical Observatory, Center for Astronomy, University of Hyogo,\\
                   407--2 Nishigaichi, Sayo-cho, Hyogo 679--5313, Japan}
\email{$^{*}$takayama@nhao.jp}

\KeyWords{infrared: stars -- stars: AGB and post-AGB -- stars: late-type}

\maketitle

\begin{abstract}
30$\%$--50$\%$ of the luminous pulsating red-giant stars show light variations of a longer period than the pulsation periods.
Those periods are called long secondary periods (LSP).
There has been debated  for many years but the origin of the LSP is still unknown.
To explain the LSP variations, there have been many approaches in not only observations but also theoretical studies. 
However, the invariance of the length of the LSPs has been investigated little.
Thus, we studied the temporal variations of the period by performing the weighted wavelet-Z-transform analysis.

Using the OGLE-III database, the $I$-band light curves of 6904 and 1945 LSP candidates in the Large/Small Magellanic Clouds, respectively,  were analyzed.
Most of our sample stars indicated that the period corresponding to the LSP was constant during the observation term. 
However, 101 and 44 LSP stars in the LMC and SMC, respectively, showed the signature of the temporal variations of the LSPs.
There were diversities of the period modulations i.e. monotonic increase or decrease, or constant until the middle and then increase, etc.
The comet-like companion is one of the possible explanations for the LSP variations, but this hypothesis cannot explain the period modulations because the LSP is determined by the orbital period.
\end{abstract}

\section{Introduction}

Thanks to long-term ground-based observations such as MACHO (Alcock et al. 1992) and Optical Gravitational Lensing Experiment (OGLE, Udalski et al. 1992), many of the luminous red giants in an evolutionary phase of red-giant branch (RGB) and asymptotic-giant branch (AGB) stars have been known as variable stars.
They are called long period variables (LPVs).
The pulsating stars are one of the most typical classes of the LPV. 

It has also been known that there are period-luminosity (PL) relations between the pulsation periods and stellar luminosities by the studies of the LPVs in the Large (LMC) and Small Magellanic Clouds (SMC) (e.g. Wood et al. 1999; Soszy\'nski et al. 2004, 2011).
There are, at least, 6 PL relations attributed to the stellar pulsation periods. 
In order of increasing the length of the period, the sequences A$^{\prime}$, A, B, C$^{\prime}$, F, and C have been known (Wood et al. 1999; Soszy\'nski et al. 2004; Ita et al. 2004; Tabur et al. 2010; Soszy\'nski \& Wood 2013).
The sequences C$^{\prime}$, F, and C consist of the Miras and Semi-Regular (SR) variables while the sequences A$^{\prime}$, A, B, and a part of C$^{\prime}$ are the OGLE small amplitude red giants (OSARGs) (e.g. Soszy\'nski et al. 2004; Soszy\'nski et al. 2013).
In addition, it has been thought that each PL relation is explained by different pulsation modes (e.g. Takayama et al. 2013; Wood 2015; Trabucchi et al. 2017).
The sequence C, the PL relation corresponding to the longest period among the stellar pulsations, can be explained by the radial fundamental mode (Ita et al. 2004; Trabucchi et al. 2021).

On the other hand, the light variations associated with longer periods than the sequence C have been found in such pulsating red giants.
The PL relation consisting of those periods is called the sequence D (Wood et al. 1999). 
Such longer periods are also called the Long Secondary Period (LSP).
The length of the LSPs ranges from 400 d to 1500 d which is about 4 times longer than the period in the sequence C in the same luminosity.
This indicates that the LSP variations cannot be explained by the radial pulsations on the AGB.
Hence there is ongoing debate on the origin of the LSPs.

Here, we review the hypotheses for the LSPs proposed and argued by the previous works.
The rotating star with a dark spot is one of the possible explanations for the LSPs.
The light curve shapes attributed to the LSP variations were similar to those of the known rotating spotted stars (Soszy\'nski \& Udalski 2014). 
However the rotation periods of a single star in the AGB phase would be longer than the LSPs.
Olivier and Wood (2003) found that the rotation velocities of the LSP stars were typically less than 3 km s$^{-1}$, which corresponded to the rotation period of $\sim$ 2900 d when an AGB star with R $\sim$ 170 $R_{\odot}$.
In addition, Takayama, Wood, and Ita (2015) argued that the theoretical models for an AGB star with a dark spot could not explain the observed colour and magnitude variations involved in the LSP. 

The periodic dust formation in the circumstellar space is proposed by Wood et al. (1999).
If the circumstellar dusts are formed recurrently, the light from a central star is likely  dimmed each time. 
Wood and Nicholls (2009) found that some LSP stars showed mid-infrared (IR) excess similar to the case of the R Coronae Borealis (RCrB) stars, which implied that the patchy dust clouds might surround the LSP star.
On the other hand, Takayama, Wood, and Ita (2015) investigated the colour variations associated with the LSPs among the optical and near-IR bands.
They found that the $J$ - $K_{\rm s}$ colours of the oxygen-rich LSP stars barely changed or became bluer when the star dimmed. 
They also considered the theoretical models assuming a star with a spherical dust shell in a circumstellar space.
However, the colour variations derived from the models did not agree with those of the LSP stars.

Non-radial pulsations are one of the possible explanations which have been argued for a long time (e.g. Wood et al. 1999, 2004), and  recently the potential pulsation modes were proposed by Saio et al. (2015).
They assumed the non-adiabatic and non-radial g$^{-}$-mode oscillations in a luminous AGB star.
In convective conditions, the frequencies of non-radial g$^{-}$-modes become purely imaginary in adiabatic conditions i.e. non-oscillatory (Shibahashi $\&$ Osaki 1981).
But considering strong non-adiabatic conditions, g$^{-}$-modes can be oscillatory and those oscillation modes are called oscillatory convective modes.
Saio et al. found that the oscillatory convective modes can be excited in the convective envelopes of the luminous AGB stars.
They also found that the loci of the pulsation periods along the evolutionary trucks of the AGB stars roughly agreed with the location of the sequence D on the PL diagram.
In addition, Takayama and Ita (2020) studied the light variations of the LSPs in the optical and near-IR bands.
They also assumed a non-radially oscillating star with the dipole modes and found a good agreement with the light amplitudes in the observations.

One of the most likely explanations, along with the non-radial oscillations, for the LSP variations is the comet-like companion, which was first proposed by Wood et al. (1999).
This hypothesis assumes eclipsing binaries consisting of a central star and a low-mass companion surrounded by circumstellar dust with a dusty tail (e.g. Wood et al. 1999; Soszy\'nski \& Udalski 2014).
To estimate the mass of the companion, the radial velocities have been studied.
The radial velocity amplitudes associated with the LSP variations lay mainly between 2.0 and 5.0 km s$^{-1}$ and the typical value is $\sim$ 3.5 km s$^{-1}$ (Hinkle et al. 2002; Nicholls et al. 2009). 
Assuming a binary system with a total mass of 1.5 M$_{\odot}$, the mass of the companion star was 0.09 M$_{\odot}$ (Nicholls et al. 2009).
They also argued that such binary systems are unlikely because the fraction of 1 M$_{\odot}$ main sequence stars with a 0.06--0.12~M$_{\odot}$ companion is less than 1$\%$ of the total (i.e. blown dwarf desert).
On the other hand, recently Soszy\'nski et al. (2021) studied the mid-IR photometric properties of the LSP variations and found that about half of them showed the secondary minima in their light curves.
They argued that the appearance of secondary minima in the light curves is evidence that eclipsing binary is a reasonable explanation for the origin of the LSPs.

Many explanations for the LSP phenomenon have been proposed.
Here, we show the advantages and disadvantages of each hypothesis in table \ref{tab1}.

\begin{table*}
\caption{The hypotheses proposed by previous works and the results of observations for the LSPs.}
\begin{tabular}{cccccc}
\hline
Hypothesis & \begin{tabular}{c} Periodic\\  light variations \end{tabular} & \begin{tabular}{c}  Length \\ of the LSP\end{tabular}   & PL relation & \begin{tabular}{c}Near-IR\\  colour variations\end{tabular} & \begin{tabular}{c}Secondary minima \\in the light curves\end{tabular} \\
\hline
\begin{tabular}{c}Rotating star\\ with a dark spot \end{tabular}  & Agree & Disagree & Uncertain & Disagree & Disagree\\
Circumstellar dust & Uncertain & Uncertain & Uncertain & Disagree & Disagree\\
Radial pulsations & Agree & Disagree &Uncertain & Uncertain & Uncertain   \\
\begin{tabular}{c}Non-radial \\ pulsations\\ (g$^{-}$ modes)\end{tabular}  &Agree &Agree &Agree &Agree &Uncertain\\
\begin{tabular}{c}Comet like \\ companion \end{tabular}&Agree &Agree &Uncertain &Uncertain &Agree \\
\hline
\label{tab1}
\end{tabular}
\end{table*}

As mentioned above, there have been many attempts to elucidate the origin of the LSPs from the discussion of many properties of the light curves.
However, the periodicity of the light variations in the LSP has little understanding.
The explanations for the LSP variations would be constrained depending on whether the periodicity of the LSP phenomenon is strong (i.e. regular) or weak (i.e. irregular).
For example, the eclipsing binaries including the comet-like companion likely show strong periodicities since the period of the light variations agrees with the orbital period. 
In other words, if the evidence of the period modulations is found in the light curves associated with the LSP, the eclipsing binaries are inconsistent with the LSP phenomenon.
On the other hand, the pulsation periods can change when the mean radius of the star changes.
Thus, we perform the wavelet analysis of the optical light curves of the LSP stars and investigate the length of the LSP at cycle-to-cycle. 
In this paper, we discuss the constraint against the explanations for the LSP proposed in previous works.

\begin{figure*}[h!]
 \begin{minipage}{0.45\linewidth}
  \centering
    \FigureFile(80mm,00mm){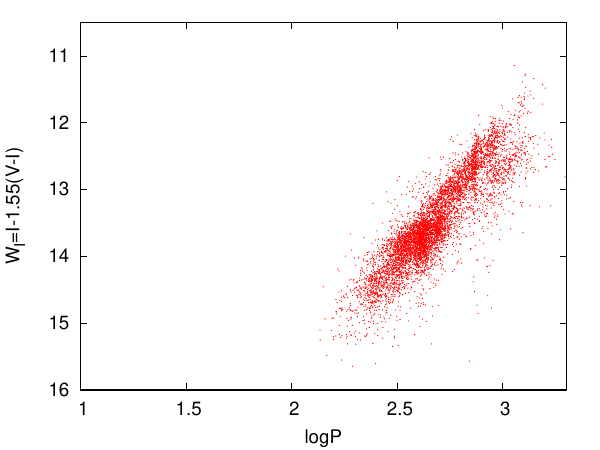}
\end{minipage}
 \begin{minipage}{0.45\linewidth}
    \FigureFile(80mm,00mm){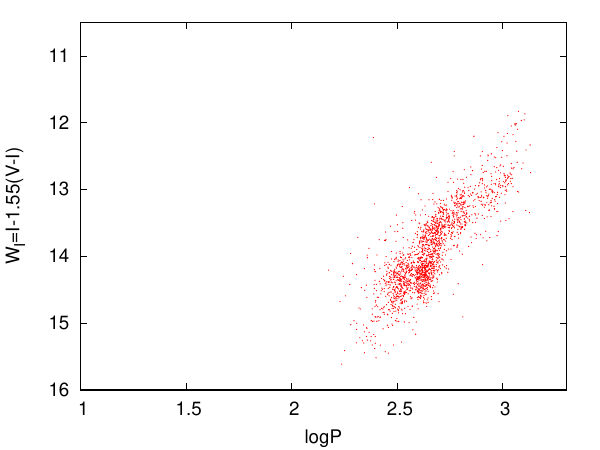}
\end{minipage}
 \caption{[Left] Period vs $W_{I}$ diagram with the sample of the LMC LSP candidates. [Right] same as the left panel but of the SMC LSP candidates.}
  \label{PL}
\end{figure*}

\section{Data}

We investigate the $I$-band time series obtained by OGLE-II $\&$ -III survey. 
OGLE-II $\&$ -III are the long-term photometric surveys from 1997 to 2000 (Udalski et al. 1997; Szyma\'nski 2005) and 2001 to 2009 (Udalski et al. 2008), respectively, with the $V$ and $I$ bands and observed each star about 500--1000 times with the $I$ band.   
Soszy\'nski et al. (2009, 2011) obtained 6951 and 1953 LSP candidates in the Large (LMC) and Small Magellanic Clouds (SMC), respectively, from the OGLE database according to the position on the PL diagram.
We selected 6904 and 1945 LSP stars in the LMC and SMC, respectively, which have been published in the time series with both the $V$ and $I$ bands within their catalogues.
Figure~\ref{PL} shows the PL diagrams with our sample of LSP stars in the LMC and SMC, respectively, where $W_{I}$ is the Wesenheit index \nobreak($W_{I} = V - 1.55(V - I)$).
We confirmed that the positions of those sample stars were consistent with the location of the sequence D on the PL diagrams.

\section{Analysis}
\label{ana}

The observation times of OGLE survey are unequally spaced.
In addition, there is annual discontinuation for several month due to its ground-based observations.
Thus we adopted the weighted wavelet-Z-transform (WWZ, Foster 1996b) with $I$-band light curves of the OGLE database.
The WWZ assumes a Morlet wavelet $F_{\tau,\: \omega}(t)$ for a given time $t$ described as below,
\begin{eqnarray}\label{morlet}
F_{\tau,\: \omega}(t) = A_{\tau,\:\omega}&+&a_{\tau,\:\omega}G_{\tau,\: \omega}(t)\sin [\omega(t-\tau)]\nonumber \\
&+&b_{\tau,\:\omega} G_{\tau,\: \omega}(t) \cos [\omega(t-\tau)],
\end{eqnarray}
where $\tau$ and $\omega$ are a program time and frequency, respectively.
$G_{\tau,\: \omega}(t)$ is a statistical weight with a constant $c$,
\begin{equation}\label{weight}
G_{\tau,\: \omega}(t)=e^{-c \omega^2 (t-\tau)^2}.
\end{equation}
As shown above, the Morlet wavelet is the Fourier wavelet with a Gaussian decay profile which means the waves localized near $t=\tau$.
The constant $c$ decides the rate of the Gaussian decay against the period $2\pi / \omega$ of the Fourier components. 
The decay span that the Gaussian decays to 1/$e$ is $1/(\sqrt{c} \omega)$.
Hence the number of the periods within the span is $1/(2\pi \sqrt{c})$.
We adopt $c=1/(8\pi^2)$ according to Foster (1996b) i.e. $\sqrt{2}$ cycles are included during the decay span.
Adapting the least square fit with the Morlet wavelet $F_{\tau,\: \omega}(t)$ for the light curves, the average value of the wavelet $A_{\tau,\:\omega}$ and the Fourier coefficients $a_{\tau,\:\omega}$ and $b_{\tau,\:\omega}$ are derived.

To investigate the light variations associated with the LSP,  the wavelet analysis is performed with a range of $\omega\: \lesssim$ 0.063 d$^{-1}$ i.e. $P \: \geqq$ 100 d.
We also adopt the range for $\tau$ from $t_{\rm start}+1/(\sqrt{c} \omega)$ to $t_{\rm end}-1/(\sqrt{c} \omega)$ so that the central part of the Morlet wavelet falls enough within the full light-curve interval, where $t_{\rm start}$ and $t_{\rm end}$ are the start and end time of the light curves, respectively.
The fit is performed with the intervals of $\delta \tau$=30 d and $\delta \log \omega$=0.01.

Considering the $Z$-statistic of Foster (1996a), we discuss the power spectrum of the WWZ.
We define the effective number $N_{\rm eff}$ according to Foster (1996b),
\begin{eqnarray}\label{Neff}
N_{\rm eff}=\frac{[\sum_{i} G_{\tau,\: \omega}(t_{i})]^{2}}{\sum_{i} [G_{\tau,\: \omega}(t_{i})]^{2}},
\end{eqnarray}
where $t_{i}$ is $i$ th observation time.
And also the weighted variation of the data and the Morlet wavelet are described as below, respectively, 
\begin{eqnarray}\label{VI}
V_{I}=\frac{\sum_{i} G_{\tau,\: \omega}(t_{i}) [I(t_{i})]^2}{\sum_{i}G_{\tau,\: \omega}(t_{i})} - \left[ \frac{\sum_{i} G_{\tau,\: \omega}(t_{i}) I(t_{i})}{\sum_{i}G_{\tau,\: \omega}(t_{i})} \right]^{2},
\end{eqnarray}
and
\begin{eqnarray}\label{VF}
V_{F}=\frac{\sum_{i} G_{\tau,\: \omega}(t_{i}) [F_{\tau,\: \omega}(t_{i})]^2}{\sum_{i}G_{\tau,\: \omega}(t_{i})} - \left[ \frac{\sum_{i} G_{\tau,\: \omega}(t_{i}) F_{\tau,\: \omega}(t_{i})}{\sum_{i}G_{\tau,\: \omega}(t_{i})} \right]^{2},\nonumber\\
\end{eqnarray}
where $I(t_{i})$ is the observation data of magnitudes.
Then the power of the WWZ at ($\tau, \: \omega$) is given as below,
\begin{eqnarray}\label{Z}
Z=\frac{(N_{\rm eff}-3)V_{F}}{2(V_{I}-V_{F})}.
\end{eqnarray}
In the next section, we report the result of the wavelet analysis with the power spectrum and discuss the temporal variations of the length of the LSP.

\begin{figure*}[h!]
 \begin{minipage}{0.5\linewidth}
  \centering
  \FigureFile(75mm, 0mm){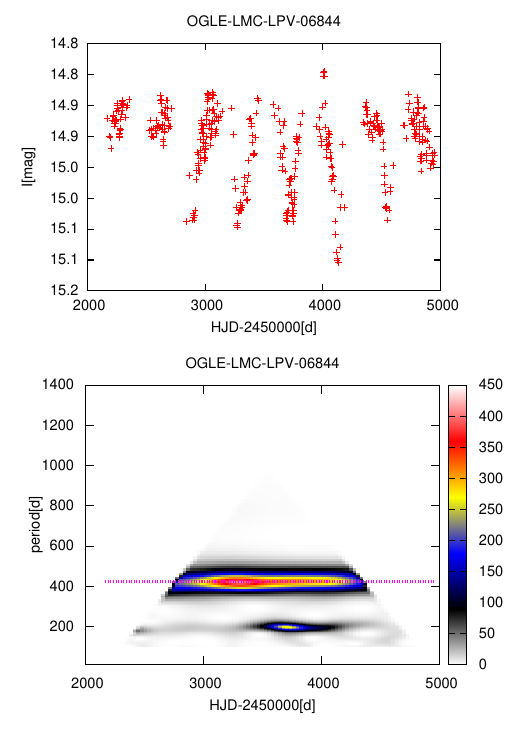}
 \end{minipage}
 \begin{minipage}{0.5\linewidth}
  \centering
  \FigureFile(75mm, 0mm){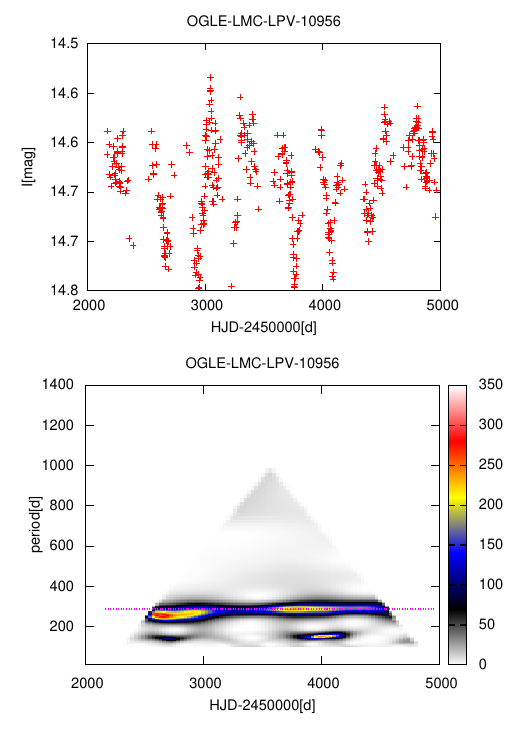}
 \end{minipage}
\caption{[Left column]The light curves (upper) and contour map (lower) of the power spectrum of OGLE-LMC-LPV-06844. The observation errors in the $I$ band on the light curves are typically smaller than about 0.01 magnitude. The colours in the contour map show the power corresponding to $Z$ of the WWZ. The horizontal dotted magenta line shows the period published by the OGLE project. [Right column]Same as the left column but of OGLE-LMC-LPV-10956.}
\label{sample_sp}
\end{figure*}

\begin{figure*}[h!]
\begin{minipage}{0.5\linewidth}
 \centering
  \FigureFile(75mm, 0mm){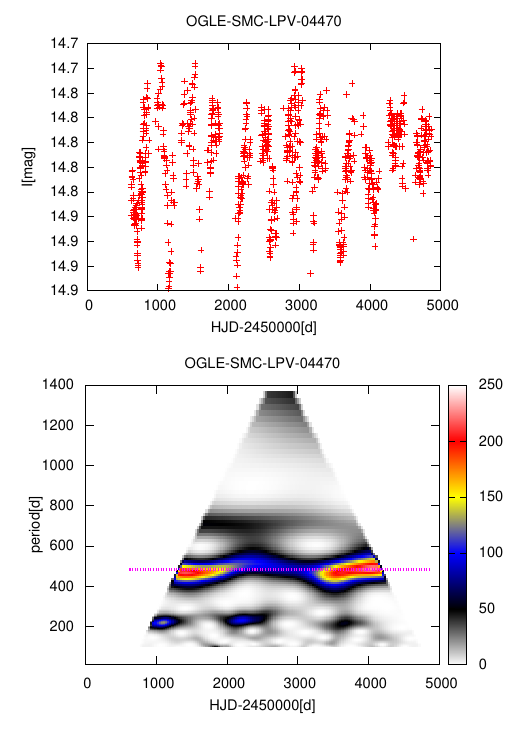}
\end{minipage}
\begin{minipage}{0.5\linewidth}
  \FigureFile(75mm, 0mm){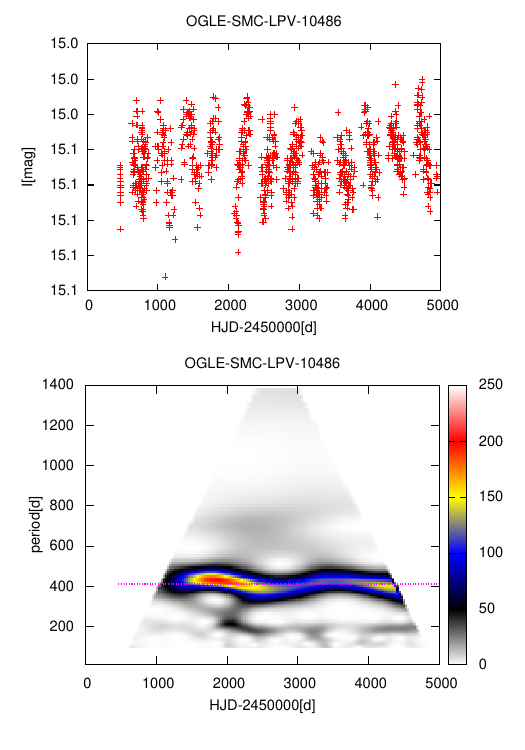}
\end{minipage}
\caption{Same as figure~\ref{sample_sp} but of [left] OGLE-SMC-LPV-04470 and [right] a sample of the LPVs showing the light variations by binarity (OGLE-SMC-LPV-10486).}
\label{meandering}
\end{figure*}

\section{Results and discussion}
\subsection{Light curves and power spectrum}
Figure~\ref{sample_sp} shows samples of the light curves and contour maps of the power spectrum derived from the WWZ.
Most of our sample LSPs showed a good agreement of the period published by the OGLE with the period corresponding to the peak of the power spectrum. 
As mentioned in section~\ref{ana}, the width of the central part of the Morlet wavelet depends on the program frequency $\omega$ (i.e. period).
Thus, the range of the program time $\tau$ which our wavelet analysis can perform becomes shorter in a longer program period.
This is the reason why the coloured region on the contour map is triangle shape.
It also means that our method can only performed when the length of the LSP is $1/(2\sqrt{2})$ times shorter than the time span of the observations.

\subsection{Case A: regular light variations}
Although there were diversities of the pattern of the contour map of both the LMC/SMC stars, the most typical  contour map shows the power spectrum in which the period corresponding to the peak is almost constant.
Two contour maps of figure~\ref{sample_sp} correspond to the examples of the contour maps with horizontal ridges.
The light curves of those samples are typically regular and show prominent light variations associated with the LSP.

As shown in the left-bottom panel of figure~\ref{sample_sp}, OGLE-LMC-LPV-06844 shows a power spectrum with a continuous ridge on the contour map. 
On the other hand, we also found that some contour maps show multi-humps or discrete ridges along the horizontal axis.
One of the explanations for such detached ridges is the amplitude variations in the light curves. 
The right column of figure~\ref{sample_sp} is the light curves and contour map of OGLE-LMC-LPV-10956.
We can see variations in the light amplitudes.
The regions corresponding to larger amplitudes among the light curves were consistent with the location of the ridges corresponding to large power.  
However, in order to consider other possible explanations, further research is needed.

Another case of the contour map was a slightly meandering ridge as shown in the left-bottom panel of figure~\ref{meandering}.
We can see a ridge meandering around the horizontal dotted line.
This might indicate period variations of LSP but it is unlikely by analogy of eclipse and ellipsoidal binaries.
We studied light curves of the red-giant binaries in the SMC obtained by Soszy\'nski et al. (2011).
The right-bottom panel of figure~\ref{meandering} shows the contour map of a sample of the red-giant binaries showing a meandering ridge.
The periods of the light variations caused by those eclipse/ellipsoidal binaries should be consistent with the orbital periods and, thus generally are invariant.
Further studies are needed to explain the reason for the meandering ridges found in the eclipse/ellipsoidal binaries.
On the other hand, at least, the amounts of the period variations larger than the meandering obtained in the eclipse/ellipsoidal binaries would be required for the evidence of the modulations of the LSPs.

\subsection{Case B: period modulation}
\begin{figure*}[p!]
\begin{tabular}{cc}
 \begin{minipage}{0.5\linewidth}
 \centering
  \FigureFile(75mm, 0mm){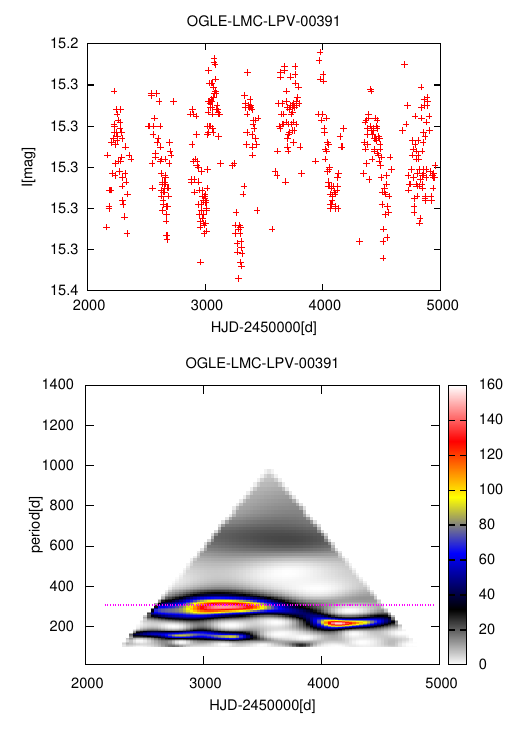}
 \end{minipage}

\begin{minipage}{0.5\linewidth}
 \centering
  \FigureFile(75mm, 0mm){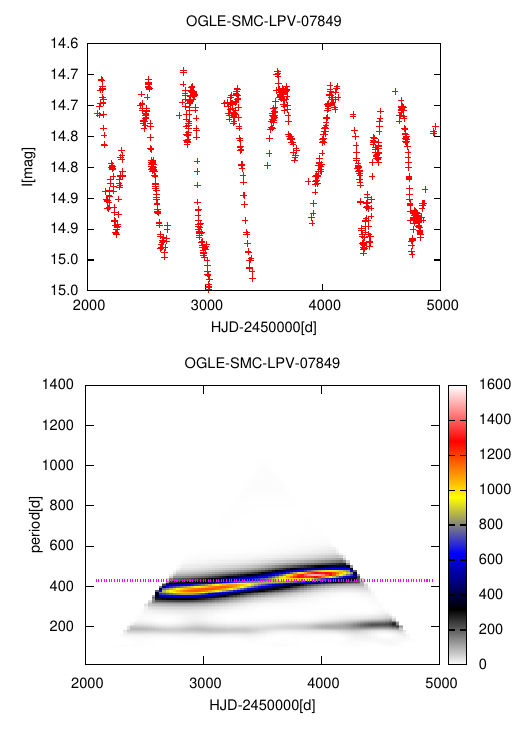}
 \end{minipage}\\

 \begin{minipage}{0.5\linewidth}
 \centering
  \FigureFile(75mm, 0mm){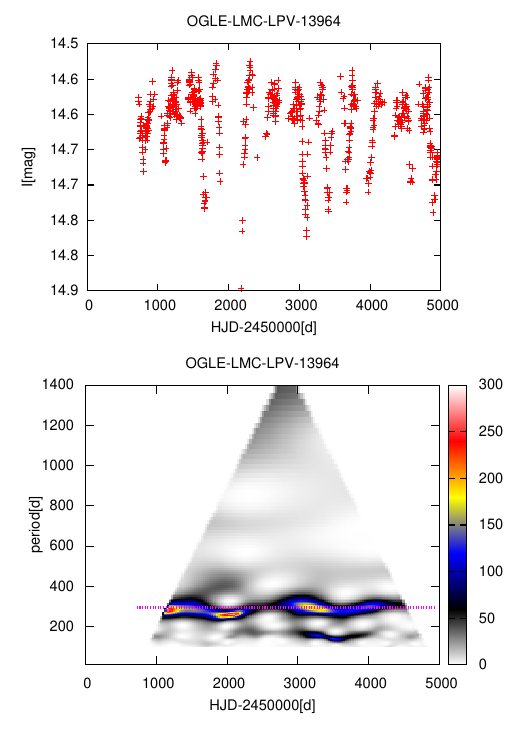}
 \end{minipage}

 \begin{minipage}{0.5\linewidth}
 \centering
  \FigureFile(75mm, 0mm){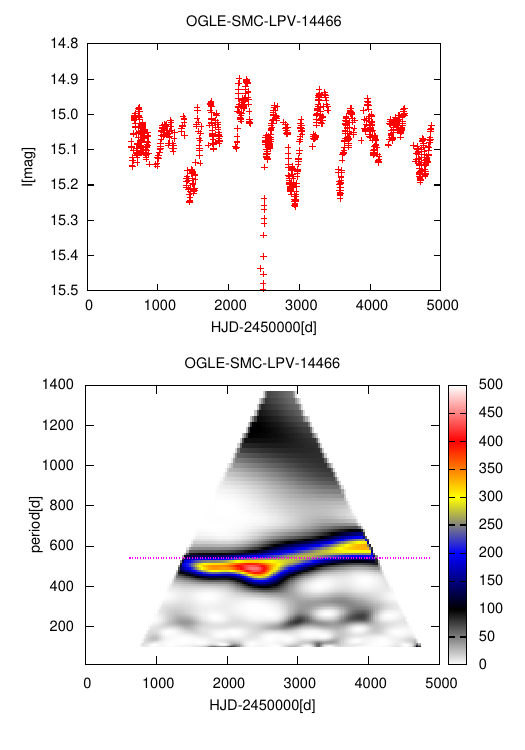}
 \end{minipage}
 
\end{tabular}
\caption{Same as figure~\ref{sample_sp} but of [left-top] OGLE-LMC-LPV-00391, [right-top] OGLE-SMC-LPV-07849, [left-bottom] OGLE-LMC-LPV-13964, and [right-bottom] OGLE-SMC-LPV-14466, respectively.}
\label{Pvari}
\end{figure*}

\begin{figure*}[h!]
\begin{minipage}{0.5\linewidth}
 \centering
  \FigureFile(70mm, 0mm){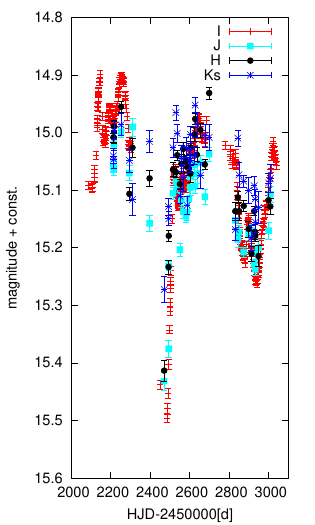}
\end{minipage}
\begin{minipage}{0.5\linewidth}
  \FigureFile(70mm, 0mm){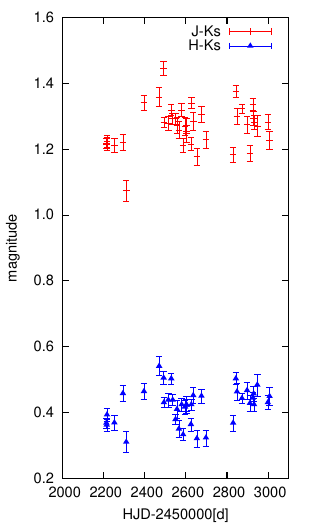}
\end{minipage}
\caption{[Left] the time series of the magnitudes in the $I$, $J$, $H$, and $K_{\rm s}$ bands of OGLE-SMC-LPV-14466. [Right] the time series of the $J-K_{\rm s}$ and $H-K_{\rm s}$ colours of OGLE-SMC-LPV-14466.}
\label{IJHK14466}
\end{figure*}

\begin{figure}[h!]
 \centering
  \FigureFile(85mm, 0mm){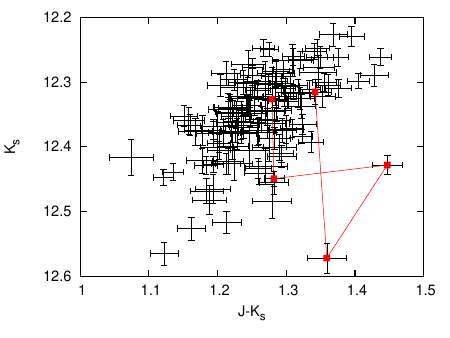}
\caption{The $J-K_{\rm s}$ vs $K_{\rm s}$ diagram plotted with OGLE-SMC-LPV-14466. The black filled circles are the locus of the observations from HJD = 2,452,212.5 d to HJD = 2,458,083.3 d with SIRIUS camera. The red solid lines correspond to the locus near the large dimming (HJD = 2,452,395.6--2,452,514.4).}
\label{CMD}
\end{figure}

As shown above, we studied the light curves and power spectrum of all our samples by eyes.
The power spectrum of most of our samples suggested that the period associated with the LSPs was constant or at least barely changed during the observation term.
Although some of those spectra showed fluctuation of the power, the period corresponding to the peak on the contour map was almost invariant.
 
However, some power spectra seemed to indicate a strong temporal change according to the contour map. 
This implies the period modulations.
The features of those ridges on the contour maps were diverse and complex so it was difficult to give criteria in a quantitative way. 
Hence we set a simple rule to determine by eyes the candidates showing the period modulations as below.
 \begin{itemize}
  \item[] {\it The periods corresponding to the LSP have changed as similar to or larger than about the full-width of half maximum of the ridge on the contour map even once.}
 \end{itemize}

According to the above criteria, we found that the sample of 101 and 44 LSPs in the LMC and SMC, respectively, showed significant changes in the period corresponding to the peak of the power spectrum.
The ratios of the number of possible candidates against the whole of our sample were 1.5$\%$ and 2.3$\%$ in the LMC and SMC, respectively.

The temporal changes of the periods corresponding to the LSP variations would give us a strong limitation for the explanations for the LSP.
In the comet-like companion, the length of the LSP should be consistent with the orbital period which would be constant.
This hypothesis needs to reconsider how to explain the change in the orbital period.

On the other hand, the pulsation periods in luminous AGB stars have theoretically been predicted to change the length during the cycle of the thermal pulse on the AGB star.
Vassiliadis and Wood (1993) considered Mira-type stars in the evolution phase of thermally pulsating AGB (TP-AGB) stars.
Once a thermal pulse occurs, the stellar surface luminosities rapidly increase and decrease like a ``spike", then slowly recover for a long time.
The pulsation periods corresponding to the radial fundamental modes also showed rapid change in a short time and slow recovery.
Those results indicate that the effective temperatures and stellar radii also change depending on the surface luminosities.

The intervals of the cycle-to-cycle of the thermal pulse were 10$^{3}$--10$^{5}$ yr, depending on the stellar masses and the chemical compositions, while the spike was only $\sim$ 500~yr.
Thus, the ratios of the timescale corresponding to the short burst and the quiescent phase are $\sim$ 10$^{-3}$--10$^{-1}$.
As mentioned above, the proportions of the LSP stars showing the period modulations were 1.5$\%$ and 2.3$\%$ in the LMC and SMC, respectively.
If the origin of the LSP is stellar pulsations, the LSP stars showing the period modulations might be interpreted as the thermal pulse occurring.
Further studies need to conclude. 


Figure~\ref{Pvari} shows the example of the power spectrum with prominent temporal variations suggesting period modulations in the LSP variations.
In most cases, the periods corresponding to the peak of the power spectrum monotonically increased or decreased during the observation term.

On the other hand, the contour map of OGLE-LMC-LPV-13964 showed the meandering ridge as similar to OGLE-SMC-LPV-04470 shown in figure~\ref{meandering} .
However the ridge corresponding to the pulsation periods of 170.82 d and 152.05 d published by OGLE also meandered in synchrony with the ridge corresponding to the LSPs.
This might indicate that the length of the LSPs correlates with the stellar radii since the pulsation periods depend on the stellar radii and masses (i.e. the mean densities of a star).
Further study is needed for the correlations between the pulsation periods and the LSPs.

Another notable object is OGLE-SMC-LPV-14466.
Only OGLE-SMC-LPV-14466 showed that the periods corresponding to the LSP variations suddenly started to increase. 
The right-bottom panel of figure~\ref{Pvari} shows the light curves and power spectrum obtained from OGLE-SMC-LPV-14466.
As you can see, the ridge corresponding to the LSPs is partly  horizontal on the contour map, which indicates that the periods of the LSP variations were almost constant until near the HJD of 2,452,500. 
On the other hand, during the later term of the observations, the ridge suggests a continuous increase in the length of the LSP. 

We also notice that according to the light curves, the large dimming occurred in OGLE-SMC-LPV-14466 near HJD of 2,452,480. 
To study the properties of the large dimming, we combined the OGLE $I$-band magnitudes with the light variations in the near-IR bands obtained with the SIRIUS camera in the IRSF 1.4 m telescope (Ita et al. 2018).
The left panel of figure~\ref{IJHK14466} shows the $I$-, $J$-, $H$-, and $K_{\rm s}$-band magnitudes near the large dimming in the $I$ band.
 As you can see, the feature of the large dimming has been obtained from not only the $I$-band light curves but also the near-IR band light curves.  
The right panel of figure~\ref{IJHK14466} shows the colour variations of the $J-K_{\rm s}$ and $H-K_{\rm s}$.
Although the light curves showed considerable dimming in both the optical and near-IR bands, the colours at the time were not noticeably redder than the colours at the normal maxima.
We also show the locus of OGLE-SMC-LPV-14466 on the $J-K_{\rm s}$ vs $K_{\rm s}$ diagram obtained by the SIRIUS camera in figure~\ref{CMD}. 
The locus corresponding to the large dimming disagreed with the locus of the normal LSP variations.
Those results would indicate that the large dimming was unlikely to be involved by the LSP variations of the star. 
However, the power spectrum obtained by the WWZ suggests that just after the large dimming occurred, the LSP started to increase.
Thus, the trigger of both the increase in the length of the LSP and the large dimming might be the same physics.

\section{Summary}
We investigated the light curves of 6904 and 1945 luminous red-giant variables showing a prominent LSP in the LMC and SMC, respectively.
By using the WWZ, we studied the temporal variations of the periods corresponding to the LSPs.

Most power spectrum of our sample of the LSP stars indicated that the period corresponding to the LSPs would be invariant.
However, we found that the power spectrum of 101 and 44 LSPs in the LMC and SMC, respectively, showed the signature of the period modulations of the LSPs.
There were diversities of the period modulations i.e. monotonic increase or decrease, or constant until the middle and then increase, etc.
The comet-like companion is one of the possible explanations, but this hypothesis cannot explain the period modulations because the LSP is determined by the orbital period.

The proportions of the LSP stars showing the period modulations against the whole of our samples in the LMC and SMC were 1.5$\%$ and 2.3$\%$, respectively.
Considering TP-AGB stars, the luminosity burst due to the thermal pulse accompanies  rapid change of the pulsation periods and it would end in $\sim$ 500 yr, while the quiescent phase after the burst would continue for 10$^{3}$--10$^{5}$ yr  (Vassiliadis $\&$ Wood 1993).
Thus, it can be estimated that the AGB stars in the progress of the luminosity burst at a certain moment are about 0.001$\%$--0.1$\%$ of the whole TP-AGB stars.
If the LSP variations are associated with the stellar pulsation as proposed by Saio et al. (2015), the period modulations found in this work might be interpreted as the thermal pulse in the AGB stars.
Further studies needed for the reasonable explanation of the LSPs.


\end{document}